\documentclass[aps,prl,showpacs,twocolumn,nofootinbib,amsmath,amssymb,floatfix,altaffilletter,superscriptaddress,10pt,preprintnumbers]{revtex4-2}
\usepackage[font=small,labelfont=bf,justification=justified,format=plain]{caption}
\usepackage{graphicx}
\usepackage{xcolor}
\usepackage{natbib}
\usepackage{mathtools}
\usepackage{physics}
\usepackage[thinc]{esdiff}
\usepackage{soul}
\usepackage{makecell}
\usepackage{enumitem}

\usepackage{ragged2e}
\usepackage{hyperref}
\hypersetup{
    linktocpage=true,
    colorlinks=true,
    citecolor=black!30!blue,
    urlcolor=black!30!blue,
    linkcolor=black!30!blue,
    breaklinks=true,
    }
\renewcommand{\L}{\mathcal{L}}
\newcommand{\K}{\mathcal{K}}

\newcommand{\C}{\mathcal{C}}
\renewcommand{\d}{\text{d}}

\begin{document}

\title{$\K$-nonizing}
\author{Jose Beltr{\'a}n Jim{\'e}nez}
\email{jose.beltran@usal.es}
\affiliation{Departamento de Física Fundamental and IUFFyM, Universidad de Salamanca, E-37008 Salamanca, Spain}
\author{Teodor Borislavov Vasilev}
\email{teodorbo@ucm.es}
\author{Dar\'{\i}o Jaramillo-Garrido}
\email{djaramil@ucm.es}
\author{Antonio L. Maroto}
\email{maroto@ucm.es}
\author{Prado Martín-Moruno}
\email{pradomm@ucm.es}
\affiliation{Departamento de F\'isica Te\'orica and Instituto de F\'isica de Part\'iculas y del Cosmos (IPARCOS-UCM),
Universidad Complutense de Madrid, 28040 Madrid, Spain}

\begin{abstract} 
We unveil the dynamical equivalence of field theories with non-canonical kinetic terms and canonical theories with a volume element invariant under transverse diffeomorphisms.
The proof of the equivalence also reveals a subtle connection between the standard Legendre transformation and the so-called Clairaut equation. Explicit examples of canonizable theories include classes of $k$-essence, non-linear electrodynamics, or $f(R)$ theories.
The equivalence can also be extended to the class of mimetic theories.
\end{abstract}

\preprint{IPARCOS-UCM-25-044}
\maketitle

\noindent\textbf{\textit{Introduction.---}} 
Although it is probably in the context of gravitation where they have been more extensively used in recent years, field theories with non-canonical kinetic terms arise in different areas of physics ranging from hadron physics \cite{Weinberg:1978kz} to string theory \cite{Green:2012pqa}, going through condensed matter  \cite{Shapere:2012nq,Wilczek:2012jt}, classical and quantum electrodynamics \cite{Born:1933pep,Born:1934gh,Heisenberg:1936nmg}, and the description of topological defects \cite{Skyrme:1961vq,Babichev:2006cy}. In gravitation the paradigmatic examples are \textit{k}-essence and $f(R)$-gravity, which find applications in cosmic inflation \cite{Armendariz-Picon:1999hyi,Garriga:1999vw,DeFelice:2010aj}, dark energy \cite{Armendariz-Picon:2000nqq,Scherrer:2004au,Sotiriou:2008rp,Nojiri:2010wj}, and black hole physics \cite{Graham:2014mda,Graham:2014ina,Sotiriou:2011dz,delaCruz-Dombriz:2009pzc}.

Non-canonical theories can appear as effective descriptions of a more fundamental underlying theory and provide a very useful tool to study the behavior of the relevant low-energy degrees of freedom. However, in many cases, their intrinsic nonlinearity hinders a proper study of their dynamics both at the classical and quantum levels. Thus, whereas canonical kinetic terms lead to Hamiltonians quadratic in momenta, through well-defined bijective transformations, the corresponding transformations for non-canonical fields can be more subtle, spoiling in many cases the standard canonical quantization procedure \cite{Wilczek:2012jt}. 
In order to overcome these difficulties, field redefinitions and the introduction of new auxiliary fields are in general needed to reshape the theory into a canonical form with new interaction terms. These \textit{canonizing} procedures \cite{Malquarti:2003nn,Aguirregabiria:2004rc,dePutter:2007ny,Babichev:2018twg} allow to translate well-established results from canonical theories into the equivalent non-canonical model. Thus, for instance, in \cite{Malquarti:2003nn}  it was shown that  product \textit{k}-essence theories can be generally rewritten as canonical theories with two scalar fields interacting through a multiplicative coupling, thus establishing an equivalence between \textit{k}-essence and the more standard quintessence theories.

In this Letter we make the observation that as the nonlinear couplings can  be interpreted 
in certain cases as a modification of the volume element in the action,  
a redefinition of the integration measure can {\it canonize} the action without the need of introducing new fields.  This will allow us to show  that 
certain classes of non-canonical field theories are dynamically equivalent to 
canonical theories, but with a reduced invariance under diffeomorphisms with unit Jacobian. These restricted 
transformations belong precisely  to the subgroup of  transverse diffeomorphisms (TDiff) \cite{Maroto:2023toq,Jaramillo-Garrido:2023cor,Alonso-Lopez:2023hkx,Jaramillo-Garrido:2024tdv,Maroto:2024mkx,Maroto:2024roe,Tessainer:2024ewm,deCruzPerez:2025ytd}.


\vspace{5pt}
\noindent\textbf{\textit{Canonizing a single Lagrangian.---}} Let $\L=\L(\Phi_A,\partial\Phi_A, \partial\partial\Phi_A,\hdots)$ be a {\it canonical} Lagrangian for arbitrary fields $\Phi_A$. Given a diffeomorphism (Diff) invariant non-canonical theory 
\begin{equation}\label{eq: diff K(L)}
    S_{\text{Diff}} = \int\text{d}^4x \sqrt{g}\, \K(\L) ,
\end{equation}
where $\K$ is an arbitrary differentiable function, we shall show that it is possible to perform a geometrical canonization by
considering a modified volume element. In other words, we will establish an equivalence with the following TDiff invariant action
\begin{equation}\label{eq: TDiff f(g) L}
    S_{\text{TDiff}} = \int\text{d}^4x \, f(g) \, \L,
\end{equation}
where the coupling $f(g)$ is some function of the metric determinant $g= \abs{\det g_{\mu\nu}}$ which must be found. We shall refer to the TDiff theory as the \textit{canonized} version of the original.

For simplicity, we can equivalently work in the covariantized formulation of \eqref{eq: TDiff f(g) L}, given by \cite{Henneaux,Jaramillo-Garrido:2024tdv}
\begin{equation}\label{eq: TDiff H(Y) L}
S_{\text{TDiff}} = \int\text{d}^4x \sqrt{g} \, H(Y)\, \L ,
\end{equation}
where the symmetry under diffeomorphisms is explicitly restored via the introduction of a Stueckelberg vector field $A^\mu$, which appears through the scalar quantity $Y=\nabla_\mu A^\mu$. The new coupling function is related to the old one through $H(Y) = Y f(Y^{-2})$.
We remark that one can always translate the results back into the manifestly TDiff formulation of the theory by evaluating the expressions in the so-called \textit{TDiff frame}, which for our purposes boils down to the substitutions
\begin{equation}\label{H(Y) and f(g)}
    Y \rightarrow \frac{1}{\sqrt{g}} ,\qquad H(Y) \rightarrow \frac{f(g)}{\sqrt{g}} ,
\end{equation}
see \cite{Jaramillo-Garrido:2024tdv} for further details. Note that in \cite{Jaramillo-Garrido:2024tdv} we demonstrated that it is possible to restore the full Diff symmetry of a TDiff theory introducing a Stueckelberg-like field, whereas in the present work we shall explore the equivalence between non-canonical Diff theories and canonical TDiff theories.

To implement the canonizing procedure, 
we shall explore if there is an \textit{on-shell} (i.e. \textit{dynamical}) equivalence between both theories.
To this end, we will compare their equations of motion (EoM), and require that they take the same form. Taking variations of the Diff invariant theory \eqref{eq: diff K(L)} we have
\begin{equation}\label{eq: delta S Diff}
    \delta S_\text{Diff} = \int \d^4x\, \delta\left[\sqrt{g}\,\K(\L) \right] .
\end{equation}
Consider now variations of the TDiff action \eqref{eq: TDiff H(Y) L}, which after discarding boundary terms read
\begin{equation}\label{eq: delta S TDiff}
    \begin{split}
        \delta S_\text{TDiff} &= \int \d^4x \bigg[ (\delta\sqrt{g}) (H - YH')\L + \sqrt{g} \, H \,\delta\L  \\
        &- \delta(\sqrt{g} \,  A^\mu) \partial_\mu\left( H' \L\right) \bigg] ,
    \end{split}
\end{equation}
where a prime denotes derivative with respect to the argument.
Let us focus on the lower line of \eqref{eq: delta S TDiff}.  The EoM for the Stueckelberg vector field $A^\mu$ give us the TDiff constraint
\begin{equation}\label{eq: constraint for K(L)}
    \nabla_\mu\left(H' \L\right) = 0 \;\Rightarrow \; H'(Y) \L = \text{const.} \equiv \lambda_0 ,
\end{equation}
implying that the last term in \eqref{eq: delta S TDiff} vanishes on-shell. Furthermore, it is easy to check that the remaining variation can be written on-shell as
\begin{equation}
        \delta S_\text{TDiff} = \int \d^4x\, \delta\Big[\sqrt{g} \big(H - YH'\big)\L\Big],
\end{equation}
so, in view of equation \eqref{eq: delta S Diff}, we can identify 
\begin{equation}
    \K(\L) = \big[H(Y)-Y H'(Y)\big] \L , \label{subeq: equivalence 1}
\end{equation}
where the constraint \eqref{eq: constraint for K(L)} must provide a well-defined on-shell relation between $\L$ and $Y$, such that one can be expressed in terms of the other.
Varying the above expression and using that $\delta(H'\L)=0$, we get $ \delta\K=H\, \delta\L$ on the constraint. In particular, this means that
\begin{equation}
    \K'(\L) = H(Y)  \label{subeq: equivalence 2}
\end{equation}
evaluated on the constraint. There are now two equivalent routes to find $H(Y)$ given $\K(\L)$.

\vspace{3pt}
\noindent $\bullet$ \textit{Legendre's route.---} Combining  \eqref{eq: constraint for K(L)}, \eqref{subeq: equivalence 1}, and \eqref{subeq: equivalence 2} yields
\begin{equation}
    \lambda_0 Y =  \L\,\K'(\L) - \K(\L) \equiv \K^\star(\L) ,
\end{equation}
where $\K^\star$ stands for the Legendre transform. If $(\K^\star)^{-1}$ is well-defined, we may invert the above to find
\begin{equation}
    \L = (\K^\star)^{-1} (\lambda_0 Y) ,
\end{equation}
which upon substitution in \eqref{subeq: equivalence 2} reveals that
\begin{equation}
    H(Y) = \K'\big[ (\K^\star)^{-1} (\lambda_0 Y) \big]
\end{equation}
is the sought-for coupling function.

\vspace{3pt}
\noindent$\bullet$ \textit{Clairaut's route.---} Using \eqref{eq: constraint for K(L)} to solve for $\L=\lambda_0/H'(Y)$, equation \eqref{subeq: equivalence 1} may be simplified to
\begin{equation}\label{eq: clairaut}
    H = YH' + G(H') ,
\end{equation}
where we have introduced the function
\begin{equation}
    G(H') = \frac{\K(\lambda_0/H')}{\lambda_0/H'} .
\end{equation}
Equation \eqref{eq: clairaut} is known as \textit{Clairaut's equation} (see e.g. \cite{arnoldGeometricalMethodsTheory1988,tenenbaumOrdinaryDifferentialEquations1985}). The general solution is the family of straight lines $H(Y) = \alpha Y + G(\alpha)$ with parameter $\alpha$. However, these linear models degenerate the constraint \eqref{eq: constraint for K(L)}, so we have to look beyond them. Indeed, we must focus on the so-called \textit{singular solution} to the equation: it is the envelope of all the straight lines and is given parametrically by
\begin{equation}\label{eq: parametric clairaut}
    \begin{cases}
        \, Y(\alpha) = -G'(\alpha) = \frac{1}{\alpha}\K'(\lambda_0/\alpha) - \frac{1}{\lambda_0}\K(\lambda_0/\alpha) ,\\[5pt]
        \, H(\alpha) = G(\alpha) - \alpha G'(\alpha) = \K'(\lambda_0/\alpha) ,
    \end{cases}
\end{equation}
with parameter $\alpha$. In order to find $H=H(Y)$, one must invert $Y = -G'(\alpha)$ to obtain $\alpha$ in terms of $Y$ and then substitute back into the expression for $H$. By the implicit function theorem this will be (locally) possible in regions where $G''(\alpha) \neq 0$, see Figure \ref{fig:Hk BI} for an illustrative example. The Clairaut route yields the same results as the Legendre one. {In fact, this Legendre--Clairaut relation also finds interesting applications in the study of constrained Hamiltonian systems (see e.g. \cite{Duplij:2010fq, Walker:2014yla, Lavrov:2016ukc} and references therein).

Finally, we stress that although we have worked in the covariantized formulation it is immediate to translate it to the manifestly TDiff theory via \eqref{H(Y) and f(g)}, thus showing that a non-canonical theory can be canonized geometrically with a suitable TDiff coupling function $f(g)$.

\begin{figure}
\includegraphics[width=\linewidth]{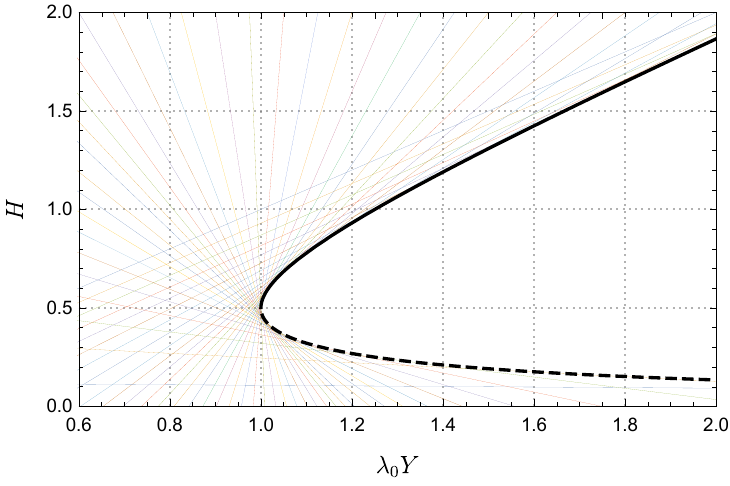}
\caption{\justifying Solutions of Clairaut's equation for the Born-Infeld model $K(X) = -\sqrt{1 - X}$. The straight line solutions generate the envelope (singular) solution $H(Y)$. The solid black line denotes the branch $X\geq 0$ of the singular solution, whereas the $X<0$ branch is shown in dashed black.}
\label{fig:Hk BI}
\end{figure}

\vspace{3pt}
\noindent $\bullet$ {\textit{Examples.---}}Let us illustrate the procedure with some examples admitting a complete treatment.

\vspace{3pt}
\noindent\textit{-- Power-law corrections.---}
Consider the class of non-canonical models of the form
\begin{equation}\label{eq: power-law corrections}
    \K(\L) = \alpha \L + \beta \L^n .
\end{equation}
They may be canonized using the TDiff coupling function
\begin{equation}
    H(Y) = \alpha + \hat\beta \, Y^\frac{n-1}{n} ,
\end{equation}
or equivalently
\begin{equation}
    f(g) = \sqrt{g} \left[\alpha + \hat\beta \, g^\frac{1-n}{2n} \right] ,
\end{equation}
with $\hat\beta \equiv (\pm1)^{n-1} n \,\beta^{\frac 1 n} \left[ \lambda_0/(n-1)\right]^\frac{n-1}{n}$, where the $\pm$ accounts for the possible branches which appear when $n$ is an even power. 

Non-canonical theories of the form \eqref{eq: power-law corrections} include, e.g., power-law \textit{k}-essences\footnote{We implicitly assume that an Einstein-Hilbert term $S_\text{EH}$ has been added to the action. Since $S_\text{EH}$ is a canonical addition, it does not affect the results obtained for $f(g)$ (see next footnote).} $K(X)=\beta X^n$ (with $X=\frac{1}{2}g^{\mu\nu}\partial_\mu\phi\partial_\nu\phi$), the interesting models $K(X) = -X + \beta X^2$ \cite{Armendariz-Picon:1999hyi,Chiba:1999ka,Scherrer:2004au,Arkani-Hamed:2003pdi}, or simple $f(R)$ theories like $f(R) = R + \beta R^n$ (which reduce to the Starobinsky model of inflation for $n=2$ \cite{Starobinsky:1980te,DeFelice:2010aj}).

\vspace{3pt}
\noindent\textit{-- Born-Infeld.---} The non-canonical models
\begin{equation}\label{eq: BI}
    \K(\L) = \alpha + \beta \sqrt{1-\L}
\end{equation}
are canonized through (note the two branches)
\begin{equation}
    H(Y) = - \frac{\beta}{2} \left[ -2 + y^2 \mp y \sqrt{y^2-4} \right]^{-1/2} ,
\end{equation}
where we denote $y \equiv 2(\lambda_0 Y + \alpha)/\beta$. 

These non-canonical theories originate in the Born-Infeld nonlinear electrodynamics \cite{Born:1933pep,Born:1934gh} $\K_\text{BI} =\lambda^4 ( 1 - \sqrt{1 - 2 \mathcal{Y}/\lambda^4 - \mathcal{Z}^2/\lambda^8} )$, with $\lambda$ an energy scale, $\mathcal{Y}= - F_{\mu\nu}F^{\mu\nu}/4$, and $\mathcal{Z}= - F_{\mu\nu}\tilde F^{\mu\nu}/4$. One may also find the form \eqref{eq: BI} in scalar field theories, where the \textit{k}-essence $K(X) = -\sqrt{1-X}$ represents the shift-symmetric Dirac-Born-Infeld model considered in e.g. \cite{Dvali:2010jz,Burrage:2014uwa} in the context of classicalization and screening mechanisms respectively, and provides an interesting theoretical basis for the Chaplygin gas in cosmology \cite{Bento:2002ps,Bouhmadi-Lopez:2016cja}. The Born-Infeld class of Lagrangians exhibit exceptional properties regarding causal propagation \cite{Deser:1998wv,Gibbons:2000xe,Mukohyama:2016ipl,deRham:2016ged}, enhanced symmetries \cite{Creminelli:2013xfa,Pajer:2018egx,Grall:2019qof} or linear response of point-like objects \cite{BeltranJimenez:2022hvs,BeltranJimenez:2024zmd} that admit a re-interpretation in the their canonized version with the potential to shed new light on their exceptional nature.


\vspace{5pt}
\noindent\textbf{\textit{Canonizing multiple Lagrangians.---}} We now discuss the situation with an arbitrary number of {\it canonical} Lagrangians, which we will collectively denote as $\boldsymbol{\L} \equiv \left(\L_1,\hdots,\L_n\right)$. The non-canonical Diff theory reads
\begin{equation}\label{eq: multidimensional Diff}
    S_{\text{Diff}} = \int\text{d}^4x \sqrt{g}\, \K(\boldsymbol{\L}) ,
\end{equation}
and we look for an on-shell equivalence with a canonical TDiff theory of the form
\begin{equation}\label{eq: multidimensional TDiff}
    S_\text{TDiff} = \int \d^4x\, \sqrt{g} \, \sum_{i=1}^n H_i(Y) \,\L_i ,
\end{equation}
where each Lagrangian has an associated volume element $H_i$. Proceeding analogously to the single-Lagrangian case, we obtain the constraint
\begin{equation}\label{eq: constraint multidimensional}
    \sum_{i=1}^n H_i'(Y)\, \L_i = \lambda_0 
\end{equation}
and the equivalence requirement 
\begin{equation}
     \K(\boldsymbol{\L}) = \sum_{i=1}^n (H_i - YH_i')\, \L_i , \label{subeq: equivalence 1 multidimensional}
\end{equation}
which lead to
\begin{equation}
     \frac{\partial \K(\boldsymbol{\L})}{\partial \L_i} = H_i(Y) . \label{subeq: equivalence 2 multidimensional}
\end{equation}
Note that in general it is not possible to canonize multiple Lagrangian theories with a universal volume element.
The question now is: how does one find the $H_i(Y)$? As we will see, in the multi-Lagrangian case it is not possible to answer this question uniquely. Indeed, we may rewrite \eqref{subeq: equivalence 1 multidimensional} as
\begin{equation}\label{eq: multidmensional Legendre}
    \lambda_0 Y = \left(\sum_{i=1}^n \frac{\partial \K(\boldsymbol{\L})}{\partial \L_i} \L_i \right) - \K(\boldsymbol{\L}) \equiv \K^\star(\boldsymbol{\L}) ,
\end{equation}
where $\K^\star$ is the multidimensional Legendre transform. To proceed further, let us assume that the Hessian matrix of $\K(\boldsymbol{\L})$ is non-degenerate, i.e. $\det( \partial^2 \K/ \partial\L_i \partial\L_j ) \neq 0$.\footnote{Note that if the contribution of a Lagrangian $\L_j$ is already canonical, then the corresponding coupling function is trivially $H_j(Y)=1$ as follows from \eqref{subeq: equivalence 2 multidimensional}. Therefore, we only ask for the non-degeneracy of the non-canonical Lagrangians.} Under this requirement, we will in principle be able to invert \eqref{subeq: equivalence 2 multidimensional} and find $\L_i = \L_i(\boldsymbol{H})$. This is then to be substituted in \eqref{eq: multidmensional Legendre} in order to find
\begin{equation}\label{eq: H-relation}
    \C(Y,\boldsymbol{H}) \equiv \K^\star[\boldsymbol{\L}(\boldsymbol{H})] - \lambda_0 Y = 0 .
\end{equation}
This equation can be understood as some sort of constraint among the coupling functions $H_i$: they cannot all be arbitrary, but must satisfy the above relation. On the one hand, this means that we can have multiple TDiff theories which canonize the same non-canonical one. On the other, and more practically, we see that we can always choose $(n-1)$ coupling functions and use this constraint to find the remaining one. Locally, we would obtain an expression of the form $H_a = H_a[Y,\{H_{i\neq a}(Y)\}]$. We now present some examples where a closed expression is reached.

\vspace{3pt}
\noindent $\bullet$ \textit{Sum of independent terms.---} Consider the theory
\begin{equation}\label{eq: sum of independent terms}
    \K(\boldsymbol{\L}) = \sum_{i=1}^n \K_i(\L_i) .
\end{equation}
The linearity and single-variable dependence of each function simplifies the treatment to finally reveal that the relation \eqref{eq: H-relation} among the coupling functions becomes
\begin{equation}
    \lambda_0 Y = \sum_{i=1}^n \K_i^\star\big[ (\K_i')^{-1}(H_i)\big] .
\end{equation}
Non-canonical theories with the structure of \eqref{eq: sum of independent terms} are found e.g. in  \textit{k}-essence \cite{Mukhanov:2005bu,Bose:2008ew}, $f(R)$ \textit{k}-essence \cite{Nojiri:2019dqc}, $f(R)$ Euler-Heisenberg electrodynamics \cite{Guerrero:2020uhn}, or gauge-flation\footnote{Gauge-flation is particularly interesting because it consists of the canonical kinetic term plus a non-canonical term of the form $(G_{\mu \nu} \tilde{G}^{\mu\nu})^2$. The canonical Lagrangian associated to this term is $G_{\mu \nu} \tilde{G}^{\mu\nu}$, which is topological when coupled to a Diff-invariant volume element. Under our canonization procedure, however, the coupling is to a TDiff-invariant volume element and this is the reason why it gives rise to a non-trivial dynamics. The same occurs for other theories described by non-linear functions of would-be topological invariants when coupled to Diff-invariant volume elements.} \cite{Maleknejad:2011jw}.

\vspace{3pt}
\noindent $\bullet$ \textit{Products of independent terms.---} The simple product
\begin{equation}\label{eq: simple product}
    \K(\L_1,\L_2) = \K_1(\L_1)\,\L_2 
\end{equation}
makes the relation \eqref{eq: H-relation} among coupling functions read
\begin{equation}
    \lambda_0 Y = H_1 \, \K_1^{-1}(H_2) .
\end{equation}
Theories of the form \eqref{eq: simple product} include e.g. product \textit{k}-essences, whose canonization is studied in \cite{Malquarti:2003nn}. The process is immediately extensible to general sums of products
\begin{equation}\label{eq: sum of simple products}
    \K(\boldsymbol{\L}) = \sum_{i=1}^n\sum_{j=1}^n \K_i(\L_i) \,\L_j ,
\end{equation}
a structure which may be found e.g. in scalar-tensor (and generalized scalar-tensor) theories of gravity \cite{Brans:1961sx,Fujii:2003pa,Faraoni:2004pi}.


\vspace{5pt}
\noindent\textbf{\textit{Uncanonizing and mimetic models.---}}
We have so far focused on the process of \textit{canonizing}, but of course, one may instead tackle the inverse problem: given a canonical TDiff theory \eqref{eq: multidimensional TDiff}, determining the equivalent non-canonical Diff description \eqref{eq: multidimensional Diff}. The equivalence requirements are again given by \eqref{subeq: equivalence 1 multidimensional} and \eqref{subeq: equivalence 2 multidimensional}, subject to the TDiff constraint \eqref{eq: constraint multidimensional}. Again, the most critical step in the procedure involves using the constraint, which can generally be written as $\mathcal F(Y,\boldsymbol{\L})=0$ for some function $\mathcal F$, to solve for $Y = Y(\boldsymbol{\L})$. If this can be done, one then simply substitutes in \eqref{subeq: equivalence 1 multidimensional} to find the sought-for $\K(\boldsymbol{\L})$, while \eqref{subeq: equivalence 2 multidimensional} acts as a consistency check. A necessary condition for this procedure to be well-defined is $\partial_Y \mathcal{F} \neq 0$, so that the constraint truly sets a relation between $Y$ and $\boldsymbol{\L}$ and not just among the Lagrangians. Hence, a TDiff model may only be recast as a non-canonical Diff theory of the form \eqref{eq: multidimensional Diff} by virtue of their on-shell equivalence if this condition is met. 

The degenerate case $\partial_Y \mathcal{F} = 0$ also leads to interesting models. The degeneracy of the constraint reads $\sum_i H_i'' \L_i = 0$, which under the assumption of independent Lagrangians implies that all coupling functions are linear, $H_i(Y) = a_i + b_i Y$. As a result, the TDiff constraint becomes a relation among Lagrangians $\sum_ib_i\L_i=\lambda_0$.
This condition, in fact, generalizes the framework of mimetic models like those studied for scalar \cite{Lim:2010yk,Chamseddine:2013kea,Golovnev:2013jxa,Sebastiani:2016ras}, vector \cite{Barvinsky:2013mea,Chaichian:2014qba}, $p$-form \cite{Gorji:2018okn} or non-Abelian gauge fields \cite{Gorji:2019ttx}.
As an example of degenerate TDiff models, it was recently shown in \cite{deCruzPerez:2025ytd} that a unified description of the $\Lambda$CDM dark sector can be achieved with a single TDiff scalar field precisely by using a linear coupling function $H(Y)$. This concrete application illustrates the fact that one can construct new classes of (phenomenologically viable) general mimetic theories starting from the degeneracy condition in the TDiff framework.


\vspace{5pt}
\noindent\textbf{\textit{Discussion.---}} In this Letter we have presented a canonizing procedure that transforms Diff invariant field theories with non-canonical kinetic terms into canonical TDiff theories with a deformed volume element. We have seen that the procedure is unambiguous for the single-Lagrangian case. However, a multi-Lagrangian non-canonical theory admits different canonizations. Future work on this front could explore the classification of TDiff theories into equivalence classes according to their dynamical behavior. On the other hand, the uncanonizing procedure is unambiguous for both the single- and multi-Lagrangian cases. That is, given a set of TDiff coupling functions $H_i$, we either recover a non-canonical theory $\K(\boldsymbol{\L})$ if the constraint is non-degenerate, or a generalized mimetic theory if it is. This, in turn, points to the fact that it is possible to treat non-canonical and mimetic theories under a common framework, namely, the TDiff one.

It is important to highlight that the canonical TDiff theory inherits the full phenomenology of its non-canonical Diff counterpart. 
This includes not only the background dynamics but also any associated pathologies or instabilities \cite{Hsu:2004vr,Babichev:2007dw,Sotiriou:2008rp,Creminelli:2008wc,Babichev:2016hys,BorislavovVasilev:2024loq} (see also \cite{Jaramillo-Garrido:2024tdv} for a detailed account of the sound speed in TDiff quintessence theories).
A particularly interesting domain of application for this equivalence is that of black hole physics, where non-canonical field theories have been actively studied \cite{delaCruz-Dombriz:2009pzc,Sotiriou:2011dz} for example in relation to the (non-)existence of hairy solutions 
\cite{Sotiriou:2011dz,Graham:2014mda,Graham:2014ina} and as dynamical models for regular black holes \cite{Ayon-Beato:2000mjt,Guerrero:2020uhn,Bronnikov:2022ofk}.
Another promising direction is cosmology, where $f(R)$-theories and $k$-essence models have been long employed to model both early-universe inflation and late-time expansion \cite{Armendariz-Picon:1999hyi,Garriga:1999vw,Chiba:1999ka,Bose:2008ew,DeFelice:2010aj,Chiba:1999ka,Bilic:2008yr,Bose:2008ew,Armendariz-Picon:2000ulo,Chiba:2002mw,Bertacca:2008uf,Nojiri:2010wj}, and one should also mention the prominent role in the interplay between local gravity and cosmological models that is played by screening mechanisms such as K-mouflage/Kinetic \cite{Babichev:2009ee,Brax:2012jr}, D-BIonic \cite{Burrage:2014uwa} or chameleon \cite{Khoury:2003aq,Cembranos:2005fi} mechanisms. 

Let us emphasize that the equivalence presented in this Letter provides a novel (geometrical) way of understanding all of these phenomena in terms of a deformed volume element. From a computational point of view, the particularities of each side of the equivalence might differ from model to model, but having different languages with which to tackle a given problem is undoubtedly a useful and valuable tool. For instance, through the procedure here presented we are able to trade off (self-)interactions in the matter sector for purely gravitational ones. This may be relevant for screening mechanisms like K-mouflage which rely on the scalar field's self-interactions, while in the TDiff formulation the scalar self-interactions have been removed and the screening now will be due to the deformed volume element.

A remarkable feature of the canonization presented in this Letter is its simplicity, since it only needs a deformation of the volume element. A relation between canonical theories formulated in extended geometries and non-canonical theories in standard geometries also exist in e.g. metric-affine theories \cite{BeltranJimenez:2017doy,Afonso:2017bxr}, but the geometrical modification is arguably more contrived since it requires a modification of the affine structure. Let us notice that our canonization procedure does not produce derivatives of $g$, but TDiff theories involving its derivatives have also been studied in e.g. \cite{Alvarez:2010cg,Lopez-Villarejo:2010uib,Blas:2011ac,Bello-Morales:2023btf,Bello-Morales:2023btf,Bello-Morales:2024vqk}. This is a crucial difference because the breakdown from Diffs to TDiffs generates a scalar that is dynamical if derivatives of $g$ are present, while it is non-dynamical in their absence. The latter is the case treated here, but the former would be interesting to e.g. canonize interactions with an additional dynamical scalar field. Finally, in this Letter, we have focused on a canonization based on a deformation of the external spacetime volume element. However, the presented procedure could also be extended to the deformation of other types of volume such as, for instance, the internal volume element of elastic media or the coset space volume in non-linear sigma models. We leave these considerations for future work.


 \vspace{5pt}

\begin{acknowledgments}
    \noindent\textbf{\textit{Acknowledgments.---}} JBJ acknowledges support from grants PID2021-122938NB-I00 and PID2024-158938NB-I00 funded by MICIU/AEI/10.13039/501100011033 and by “ERDF A way of making Europe”, the Project SA097P24 funded by Junta de Castilla y Le\'on and the research visit grant PRX23/00530. This work was supported by the MICIU (Spain) project PID2022-138263NB-I00 funded by MICIU/AEI/10.13039/501100011033 and by ERDF/EU.
    DJG acknowledges support from the Comunidad de Madrid under predoctoral contract PIPF-2023/TEC-29931.
\end{acknowledgments}

\bibliography{bibliography}


\end{document}